\documentclass{ws-mpla}

\usepackage[super]{cite}
\usepackage{xcolor}
\usepackage[verbose,hypertexnames=false]{hyperref}
\hypersetup{colorlinks=false,allbordercolors=blue,pdfborderstyle={/S/U/W 1}}

\usepackage{amssymb}
\usepackage{amsmath}
\usepackage{bm}
\usepackage{graphicx}
\usepackage{ascmac}
\usepackage{cancel}
\usepackage{braket}
\renewcommand{\Re}{\operatorname{Re}}
\renewcommand{\Im}{\operatorname{Im}}

\begin{document}

\markboth{Tomona Kinugawa and Tetsuo Hyodo}{Internal structure of near-threshold states using compositeness}


\title{Internal structure of near-threshold states using compositeness}

\author{Tomona Kinugawa}

\address{Nishina Center for Accelerator-Based Science, RIKEN, 
              Wako 351-0198, Japan \\
tomona.kinugawa@riken.jp}

\author{Tetsuo Hyodo}

\address{Research Center for Nuclear Physics, The University of Osaka, Ibaraki, Osaka 567-0047, Japan
}

\maketitle

\begin{history}
\received{(Day Month Year)}
\revised{(Day Month Year)}
\accepted{(Day Month Year)}
\published{(Day Month Year)}
\end{history}

\begin{abstract}
Understanding the internal structure of near-threshold states is essential for revealing the nature of exotic hadrons. Motivated by this challenge, we discuss the clustering structures of near-threshold $s$-wave eigenstates using the compositeness, which characterizes the clustering nature of the states. We show that shallow bound states usually possess cluster-dominant structures, while near-threshold narrow resonances are non-cluster-dominant. Through this study, we establish a theoretical foundation for the threshold energy rule, which has been known empirically. 
\end{abstract}


\ccode{PACS Nos.: 03.65.$-$w, 04.62.+v}

\section{Introduction}
\label{sec:intro}

Understanding the internal structure of exotic hadrons is a central topic in hadron physics. In recent years, many candidates for exotic hadrons have been discovered by experiments in the heavy-quark sector~\cite{ParticleDataGroup:2024cfk}, attracting significant attention from both theory and experiment. The structure of exotic hadrons is considered to differ from that of ordinary mesons, which are composed of a quark-antiquark pair, or baryons, which consist of three quarks. With four or more quarks, various structures can be realized, such as hadronic molecules, regarded as weakly bound states of two hadrons, and multiquark states, which are compact configurations of four or more quarks. Many studies have been devoted to clarifying the structure of exotic hadrons from these viewpoints~\cite{Hosaka:2016pey,Guo:2017jvc,Brambilla:2019esw,Hosaka:2025gcl}.  

To understand the structure of exotic hadrons from a more general perspective, we explore them in the context of clustering phenomena. The clustering states are characterized by a structure consisting of subunits, each composed of multiple fundamental particles relevant at that energy scale. These subunits retain their own degrees of freedom and interact with each other to form a bound state or resonance, rather than all the constituents forming a compact state directly. A well-known example is the $\alpha$-cluster structure in atomic nuclei, where $^{4}$He nuclei behave as subunits that interact with each other to form nuclei such as the ground state of $^{8}$Be~\cite{Wiringa:2000gb} and the $^{12}$C Hoyle state~\cite{Hoyle:1954zz}. In hadron physics, the formation of the hadronic molecules can be regarded as a clustering phenomenon of quarks~\cite{Hosaka:2025gcl}. In this way, clustering states are known to universally emerge across different hierarchies of matter~\cite{Nakamura:2025ivk}. Thus, clarifying the properties of clustering structures provides a general framework for understanding the internal structure of exotic hadrons. 

In general, the internal structure of states can be described as a superposition of possible components. In particular, we focus on the weight of the clustering component, called the compositeness. More precisely, the compositeness $X$ is defined as the probability of finding free scattering states of two particles $\ket{\bm{p}}$ in a bound state $\ket{B}$~\cite{Weinberg:1965zz,Hyodo:2013nka,vanKolck:2022lqz,Kinugawa:2024crb}:
\begin{align}
X &= \int \frac{d\bm{p}}{(2\pi)^{3}}\ |\braket{\bm{p}|B}|^{2}.
\end{align}
The remaining probability $Z = 1 - X$ is referred to as the elementarity, representing the weight of the non-clustering components. The compositeness $X$ has been utilized to quantitatively investigate the internal structure of exotic hadrons from the viewpoint of hadronic molecules. 

It is known that exotic hadrons tend to be observed near the scattering threshold, the lowest energy at which scattering occurs. Thus, the near-threshold region plays a crucial role in considering the structure of exotic hadrons. Furthermore, in nuclear physics, the Ikeda diagram indicates that $\alpha$-cluster states are observed near the thresholds~\cite{Ikeda:1968fry}. 
This empirical fact is called ``the threshold energy rule''~\cite{PTPS52.89}. In this way, the near-threshold energy region appears to be related to clustering phenomena. It is known that phenomena in this region are described by universal relations in terms of the scattering length (the low-energy universality)~\cite{Braaten:2004rn,Naidon:2016dpf}. As one of its consequences, Ref.~\cite{Hyodo:2014bda} has shown that in the limit where the binding energy approaches zero, the compositeness of such a state at the threshold is exactly unity. Based on this fact and the threshold energy rule, it has been naively expected that near-threshold exotic hadrons possess a molecular structure. Indeed, the deuteron, which is a shallow bound state, is well understood as a molecule of a proton and a neutron, with the compositeness $X$ being nearly unity~\cite{Weinberg:1965zz,Kamiya:2017hni,Matuschek:2020gqe,Li:2021cue,Song:2022yvz,Albaladejo:2022sux,Kinugawa:2022fzn,Oller:2022qau}. 

However, theoretical studies have shown that any state with a finite eigenenergy can in principle take an arbitrary value of the compositeness~\cite{Hyodo:2014bda,Kamiya:2016oao,Hanhart:2022qxq}. This implies that, even for near-threshold states, structures not dominated by clustering are always possible. Why then do near-threshold states such as the deuteron and $\alpha$-cluster nuclei predominantly exhibit clustering structures, while non-clustering states are rarely observed? In this contribution, we summarize our studies on the internal structure of near-threshold states~\cite{Kinugawa:2023fbf,Kinugawa:2024kwb}, which are motivated by this question. In section~\ref{sec:bound}, we investigate the internal structure of shallow bound states below the threshold whose binding energy is small using the compositeness. Then we calculate the compositeness of the near-threshold resonances above the threshold with small excitation energy in section~\ref{sec:resonances}. The last section is devoted to a summary.

\section{Structure of shallow bound states}
\label{sec:bound}

In this section, we investigate the internal structure of shallow bound states slightly below the threshold and provide a theoretical foundation for the threshold energy rule. To this end, we employ the framework of nonrelativistic effective field theory (EFT)~\cite{Braaten:2007nq} and introduce a model in which a bare state with $X = 0$ (a pure elementary state) is coupled to free scattering states. The details of the model are given in Ref.~\cite{Kinugawa:2023fbf}. The scattering amplitude $f(k)$ is calculated as 
\begin{align}
f(k)&=-\frac{\mu}{2\pi}\left[\frac{\frac{k^{2}}{2\mu}-\nu_{0}}{g_{0}^{2}}+\frac{\mu}{\pi^{2}}\left[\Lambda+ik\arctan\left(-\frac{\Lambda}{ik}\right)\right]\right]^{-1}.
\label{eq:amplitude}
\end{align}
Here, $g_{0}$ denotes the coupling constant, $\mu$ is the reduced mass, and $\nu_{0}$ is the energy of the bare state. The cutoff momentum $\Lambda$ is introduced in the loop integral to regularize the divergence. We keep $\Lambda$ finite to specify the typical scale of the system as discussed later. 

For a quantitative analysis of the internal structure of bound states, we calculate the compositeness. With a suitable choice of parameters, this model has a bound state, which can be interpreted as a dressed bare state generated by its coupling to the scattering states. The binding energy $B$ is determined by the pole condition of Eq.~\eqref{eq:amplitude} with $k = i\sqrt{2\mu B}\equiv i\kappa$. In this model, the compositeness $X$ can be expressed as a function of the model parameters and the binding energy~\cite{Kinugawa:2023fbf}:
\begin{align}
X &= \left[1+\frac{\pi^{2}\kappa}{g_{0}^{2}\mu^{2}}\left(\arctan\left(\frac{\Lambda}{\kappa}\right)-\frac{\frac{\Lambda}{\kappa}}{1+\left(\frac{\Lambda}{\kappa}\right)^{2}}\right)^{-1}\right]^{-1}.
\label{eq:X-bound}
\end{align}
Here, the dependence of the bare state energy $\nu_{0}$ is implicitly included in $\kappa$ through the bound state condition. 

Equation~\eqref{eq:X-bound} indicates that in the $B \to 0$ limit (i.e., $\kappa \to 0$), the compositeness approaches $X = 1$. Conversely, for $B \neq 0$, $X$ can in principle take any value depending on the model parameters. This leads us to examine the dependence of $X$ on the model parameters at fixed binding energy. In this way, we can investigate the internal structure that shallow bound states are likely to favor. To specify the shallow-bound states, we define a typical energy scale $E_{\rm typ} = \Lambda^{2}/(2\mu)$ determined by the cutoff. To examine the number of independent parameters, note that fixing the binding energy reduces the number of independent model parameters from three $(g_{0}, \nu_{0}, \Lambda)$ to two. In fact, from the bound-state condition, we obtain the following relation among the squared coupling constant $g_{0}^{2}$, the binding energy, and the other parameters:
\begin{align}
g_{0}^{2}(B;\nu_{0},\Lambda)&=\frac{\pi^{2}}{\mu}(B+\nu_{0})\left[\Lambda-\kappa\arctan\left(\frac{\Lambda}{\kappa}\right)\right]^{-1}.
\label{eq:g02}
\end{align}
In the calculation, we use dimensionless quantities normalized by $\Lambda$, so that the explicit cutoff dependence of the result is removed. As a result, the only remaining degree of freedom is reflected in the $\nu_{0}$ dependence of $X$. In principle, $\nu_{0}$ can be determined from a microscopic model beyond EFT. However, here we treat $\nu_{0}$ as a free parameter, thereby probing the model dependence of $X$. The range of $\nu_{0}$ is restricted to that allowed in the model:
\begin{align}
-B/E_{\rm typ} \leq \nu_{0}/E_{\rm typ} \leq 1.
\end{align}
The lower bound follows from the requirement that $g_{0}^{2} \geq 0$, and the upper bound corresponds to the applicable limit of the EFT determined by the cutoff $\Lambda$. 

\begin{figure}
\centering
  \includegraphics[width=0.75\textwidth]{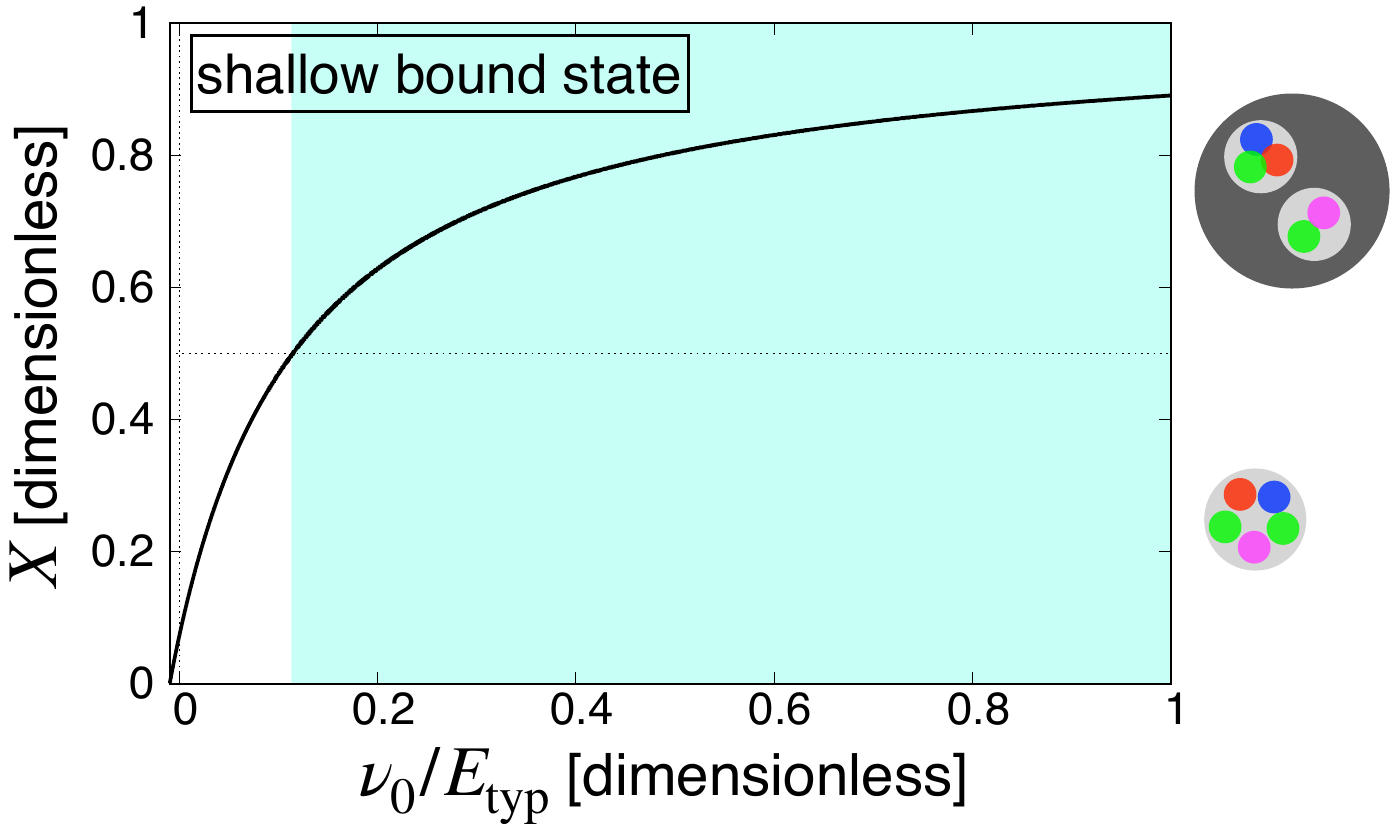}
\caption{The dependence of the compositeness $X$ on the model parameter $\nu_{0}/E_{\rm typ}$ for a shallow bound state with $B = 0.01E_{\rm typ}$.}
\label{fig:bound}      
\end{figure}

In Fig.~\ref{fig:bound}, we show the compositeness $X$ of a shallow bound state with $B = 0.01 E_{\rm typ}$ as a function of the model parameter $\nu_{0}/E_{\rm typ}$. From this plot, we find that in most of the parameter region (the colored region, 88~\%), the compositeness satisfies $X > 0.5$. This indicates that, even though the bound state originates from the bare state with $X = 0$, a shallow bound state tends to become a cluster-dominant state. This feature is explained as a consequence of the low-energy universality. At the same time, the plot shows that a non-cluster-dominant shallow bound state can also be realized if the parameter $\nu_{0}$ is tuned within the remaining 12~\% of the parameter region. However, such a case requires fine-tuning, which suggests that if the model parameters are chosen randomly, the probability of realizing a non-cluster-dominant structure is significantly lower than that of a cluster-dominant structure. These results indicate that shallow bound states naturally favor cluster dominance, thereby providing a foundation for the threshold energy rule for bound states.

\section{Structure of near-threshold resonances}
\label{sec:resonances}

In the previous section, we have demonstrated the universal nature of shallow bound states. We now turn to near-threshold resonances, which appear above the threshold with small excitation energies. To analyze the structure of such resonances, we begin with the effective range expansion (ERE), which provides the general form of the low-energy scattering amplitude~\cite{Taylor}:
\begin{align}
f(k)&=\left[-\frac{1}{a_{0}}+\frac{r_{e}}{2}k^{2}-ik\right]^{-1}.
\label{eq:ERE}
\end{align}
Here, $a_{0}$ stands for the scattering length and $r_{e}$ for the effective range. This scattering amplitude can be obtained by renormalizing Eq.~\eqref{eq:amplitude}~\cite{Braaten:2007nq}. This renormalized expression depends only on observables, and thus the compositeness $X$ can also be expressed solely in terms of observables~\cite{Hyodo:2013nka,Hyodo:2013iga,Kamiya:2016oao,Matuschek:2020gqe,Kinugawa:2022fzn,vanKolck:2022lqz,Kinugawa:2024crb}: 
\begin{align}
X&=\sqrt{\frac{1}{1-\frac{2 r_{e}}{a_{0}}}} .
\label{eq:wbr}
\end{align}
This expression of $X$ is known as the weak-binding relation.

Before discussing the compositeness of resonances, we need to introduce an interpretation scheme. This is because the compositeness of a resonance is defined as complex and cannot be interpreted as a probabilistic quantity~\cite{Hyodo:2013iga,Hyodo:2013nka,Kinugawa:2024crb}, in contrast to the case of bound states. In this work, we employ the interpretation scheme developed in Ref.~\cite{Kinugawa:2024kwb}, which defines the following three probabilistic quantities from the complex compositeness:
\begin{align}
\mathcal{X}&=\frac{(\alpha-1)|X|-\alpha|Z|+\alpha}{2\alpha-1}, \label{eq:calX} \\
\mathcal{Y}&=\frac{|X|+|Z|-1}{2\alpha-1}, \label{eq:calY} \\
\mathcal{Z}&=\frac{(\alpha-1)|Z|-\alpha|X|+\alpha}{2\alpha-1}, \label{eq:calZ}
\end{align}
where $\alpha$ is a parameter reflecting the ambiguity inherent in resonances. $\mathcal{X}$ and $\mathcal{Z}$ express the interpretable compositeness and elementarity of resonances, respectively. $\mathcal{Y}$ represents the probability of uncertain identification of the structure of resonances, which is introduced by reflecting the ambiguity of resonances discussed in Ref.~\cite{Berggren:1970wto}. 

To determine the value of $\alpha$, we focus on the region in which $\mathcal{X}$ or $\mathcal{Z}$ becomes negative and the interpretation of the internal structure of resonances breaks down~\cite{Kinugawa:2024kwb}. Because this uninterpretable region varies with $\alpha$, we choose $\alpha$ such that states with broader widths, whose internal structure is less well defined, become uninterpretable. For a quantitative criterion, we define broad resonances as states satisfying the following relation between the excitation energy $\Re E>0$ and the decay width $\Gamma$:
\begin{align}
\Re E \leq -2\,\Im E = \Gamma.
\label{eq:interpretable}
\end{align}
We assume such resonances with large decay widths have large ambiguity, and such states should be excluded from interpretation. We then choose $\alpha$ as
\begin{align}
\alpha&=\frac{\sqrt{5}-1+\sqrt{10-4\sqrt{5}}}{2}\approx 1.1318,
\end{align}
so that the structure of broad resonances is distinguished from that of narrow resonances in terms of $\mathcal{X}$, $\mathcal{Y}$, and $\mathcal{Z}$. For more details, see Ref.~\cite{Kinugawa:2024kwb}. 

\begin{figure}
\centering
  \includegraphics[width=0.75\textwidth]{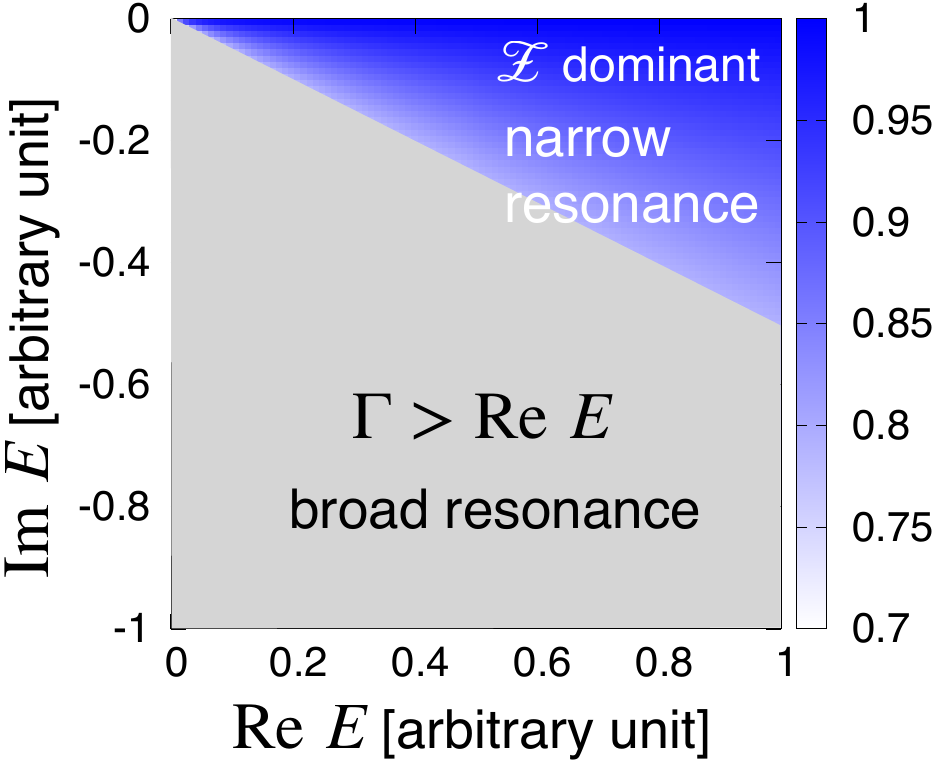}
\caption{The elementarity of resonances $\mathcal{Z}$ on the complex energy plane of the second Riemann sheet, where resonance poles exist.}
\label{fig:resonance}      
\end{figure}

Using the interpretation scheme introduced above, we investigate the internal structure of near-threshold resonances. In Fig.~\ref{fig:resonance}, we present the elementarity of resonances $\mathcal{Z}$ as a color map on the complex energy plane of the second Riemann sheet. The coordinates of the plot correspond to the eigenenergies of the resonances in arbitrary units. The gray region represents the area where the eigenenergies of broad resonances satisfying Eq.~\eqref{eq:interpretable} are located, which are excluded from the present interpretation. In the narrow-resonance region where the internal structure is interpretable in the present scheme, $\mathcal{Z}$ is close to unity; numerically, we find $\mathcal{Z} \gtrsim 0.8$. This indicates that near-threshold narrow resonances exhibit a non-cluster-dominant structure, in contrast to shallow bound states, which usually possess a cluster-dominant structure.

\section{Summary}
\label{sec:summary}

\begin{figure}
\centering
  \includegraphics[width=0.5\textwidth]{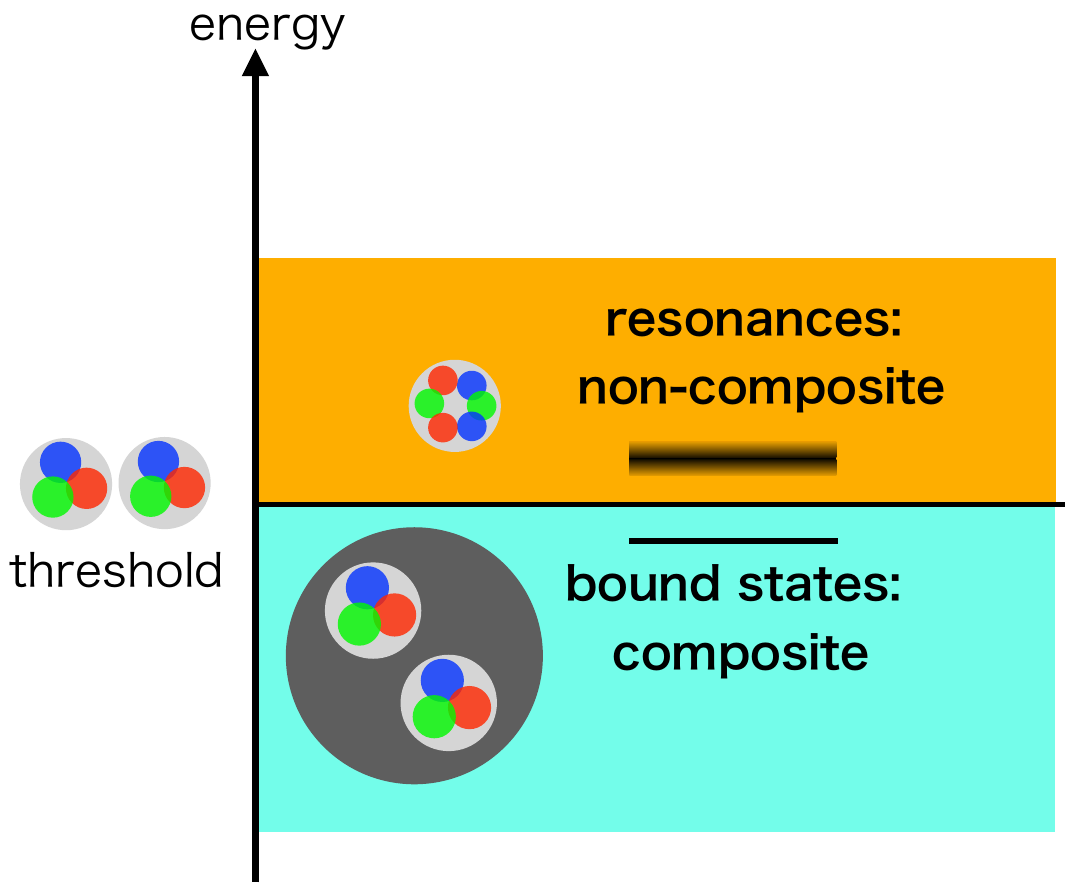}
\caption{Schematic illustration of the nature of near-threshold $s$-wave eigenstates.}
\label{fig:summary}      
\end{figure}

In this work, we examine the internal structure of near-threshold states using a quantitative measure to characterize a clustering structure, the compositeness. Within the framework of the effective field theory, we demonstrate that shallow bound states below the threshold tend to exhibit a cluster-dominant structure, thereby providing a theoretical foundation for the threshold energy rule for bound states. By introducing an interpretation scheme for complex compositeness, we also quantify the structure of above-threshold resonances. Unlike shallow bound states, near-threshold narrow resonances are found to be non-cluster-dominant. This implies that if a near-threshold composite resonance is observed, its internal structure arises from specific dynamics rather than from a universal mechanism. In this way, we have shown that the internal structures differ qualitatively depending on whether the state lies below or above the threshold (Fig.~\ref{fig:summary}).

\section*{Acknowledgments}
Tomona Kinugawa thanks the organizers of SEA-NHP 2025 for the invitation to the workshop.
This work was supported in part by the JSPS Grants-in-Aid for Scientific Research (Grant Nos.~
JP25K23387, 
JP23KJ1796, 
JP23H05439, 
JP22K03637, and 
JP18H05402) 
and by the RIKEN Special Postdoctoral Researcher Program.


\section*{ORCID}

\noindent Tomona Kinugawa - \url{https://orcid.org/0000-0002-8865-2753}

\noindent Tetsuo Hyodo - \url{https://orcid.org/0000-0002-4145-9817}


\end{document}